\documentclass[journal,10pt]{IEEEtran}

\IEEEoverridecommandlockouts

\usepackage[cmex10]{amsmath}
\usepackage[dvips]{graphicx}
\usepackage{cite}
\usepackage{amssymb}
\usepackage[eulergreek]{sansmath}
\usepackage{color}

\begin{document}

\title{Diversity Analysis of Millimeter-Wave Massive MIMO Systems}
\author{Dian-Wu Yue, Shuai Xu, and Ha H. Nguyen
\thanks{Dian-Wu Yue is with the College of Information Science and
Technology, Dalian Maritime University, Dalian, Liaoning 116026,
China (e-mail: dwyue@dlmu.edu.cn), and also with the Department of Electrical and Computer Engineering, University of Saskatchewan, 57 Campus Drive, Saskatoon, SK, Canada S7N 5A9.}
\thanks{Shuai Xu is with the College of Information Science and
Technology, Dalian Maritime University, Dalian, Liaoning 116026,
China (e-mail: xu\_shuai@dlmu.edu.cn).}
\thanks{Ha H. Nguyen is with the Department of Electrical and Computer Engineering, University of Saskatchewan, 57 Campus Drive, Saskatoon, SK, Canada S7N 5A9 (e-mail: ha.nguyen@usask.ca)}
}


\newcommand{\be}{\begin{equation}}
\newcommand{\ee}{\end{equation}}
\newcommand{\bee}{\begin{eqnarray}}
\newcommand{\eee}{\end{eqnarray}}
\newcommand{\nnb}{\nonumber}

\newcommand{\mo}{\mathbf{0}}
\newcommand{\mA}{\mathbf{A}}
\newcommand{\mB}{\mathbf{B}}
\newcommand{\mG}{\mathbf{G}}
\newcommand{\mH}{\mathbf{H}}
\newcommand{\mI}{\mathbf{I}}
\newcommand{\mR}{\mathbf{R}}
\newcommand{\mY}{\mathbf{Y}}
\newcommand{\mZ}{\mathbf{Z}}
\newcommand{\mD}{\mathbf{D}}
\newcommand{\mW}{\mathbf{W}}
\newcommand{\mF}{\mathbf{F}}
\newcommand{\mP}{\mathbf{P}}
\newcommand{\mU}{\mathbf{U}}
\newcommand{\mV}{\mathbf{V}}
\newcommand{\mSigma}{\mathbf{\Sigma}}

\newcommand{\my}{\mathbf{y}}
\newcommand{\mx}{\mathbf{x}}
\newcommand{\mz}{\mathbf{z}}

\newcommand{\mr}{\mathbf{r}}
\newcommand{\mt}{\mathbf{t}}
\newcommand{\mb}{\mathbf{b}}
\newcommand{\ma}{\mathbf{a}}

\newcommand{\mh}{\mathbf{h}}
\newcommand{\mw}{\mathbf{w}}
\newcommand{\mg}{\mathbf{g}}
\newcommand{\mf}{\mathbf{f}}
\newcommand{\mn}{\mathbf{n}}

\newcommand{\mv}{\mathbf{v}}
\newcommand{\ms}{\mathbf{s}}
\newcommand{\mmu}{\mathbf{u}}

\newcommand{\lf}{\left}
\newcommand{\ri}{\right}

\newtheorem{Lemma}{Lemma}
\newtheorem{Theorem}{Theorem}
\newtheorem{Corollary}{Corollary}
\newtheorem{Proposition}{Proposition}
\newtheorem{Example}{Example}
\newtheorem{Definition}{Definition}

\maketitle

\begin{abstract}
This paper is concerned with asymptotic diversity analysis for millimeter-wave (mmWave) massive MIMO systems. First, for a single-user mmWave system employing distributed antenna subarray architecture in which the transmitter and receiver consist of $K_t$ and $K_r$ subarrays, respectively, a diversity gain theorem is established when the numbers of antennas at subarrays go to infinity. Specifically, assuming that all subchannels have the same number of propagation paths $L$, the theorem states that by employing such a distributed antenna-subarray architecture, a diversity gain of $K_rK_tL-N_s+1$ can be achieved, where $N_s$ is the number of data streams. This result means that compared to the co-located antenna architecture, using the distributed antenna-subarray architecture can scale up the diversity gain or multiplexing gain proportionally to $K_rK_t$. The diversity gain analysis is then extended to the multiuser scenario as well as the scenario with conventional partially-connected RF structure in the literature. Simulation results obtained with the hybrid analog/digital processing corroborate the analysis results and show that the distributed subarray architecture indeed yields significantly better diversity performance than the co-located antenna architectures.
\end{abstract}


\IEEEpeerreviewmaketitle


\section{Introduction}
Recently, millimeter-wave (mmWave) communication has gained considerable attention as a candidate technology for 5G mobile communication systems and beyond \cite{Rappaport,Swindlehurst,W.Roh}. The main reason for this is the availability of vast spectrum in the mmWave band (typically 30-300 GHz) that is very attractive for high data rate communications. However, compared to communication systems operating at lower microwave frequencies (such as those currently used for 4G mobile communications), propagation loss in mmWave frequencies is much higher, in  orders of magnitude. Fortunately, given the much smaller carrier wavelengths, mmWave communication systems can make use of compact massive antenna arrays to compensate for the increased propagation loss.

Nevertheless, the large-scale antenna arrays together with high cost and large power consumption of the mixed analog/digital signal components make it difficult to equip a separate radio-frequency (RF) chain for each antenna element and perform all the signal processing in the baseband. Therefore, research on \emph{hybrid} analog-digital processing of precoder and combiner for mmWave communication systems has attracted very strong interests from both academia and industry \cite{Ayach1}$\;-$\cite{Molisch}. In particular, a large body of work has been performed to address challenges in using a limited number of RF chains for massive antenna arrays. For example, the authors in \cite{Ayach1} considered single-user precoding in mmWave massive MIMO systems and established the optimality of beam steering for both single-stream and multi-stream transmission scenarios. In \cite{Sohrabi}, the authors showed that the hybrid processing can realize any fully digital processing exactly if the number of RF chains is twice the number of data streams.

Two architectures for connecting the RF chains in the hybrid processing that have been studied in the literature are \emph{full-connected} and \emph{partially-connected}. In the former, each RF chain is connected to all the antenna elements, while only a subset of antenna elements is connected to each RF chain in the later. The partially-connected architecture is more energy-efficient and implementation-friendly since it can reduce the number of required phase shifters without significant performance loss. In the conventional partially-connected architecture \cite{Ayach2, Singh, Zhang1, He, Li} the antenna array is partitioned into a number of smaller disjoint subarrays, each of which is driven by a single transmission chain. More recently, a more general partially-connected architecture, referred to as hybridly-connected in \cite{Zhang2} and overlapped subarray-based in \cite{Song}, has been  proposed. In such a hybridly-connected structure, each sub-array is connected to multiple RF chains, and each RF chain is connected to all the antennas corresponding to the sub-array in question. In particular, the authors in \cite{Zhang2} demonstrate that the spectral efficiency of the hybridly-connected structure is better than that of the partially-connected structure and that its spectral
efficiency can approach that of the fully-connected structure with the increase in the number of RF chains.

Nevertheless, due to the facts that the antenna arrays in the above-mentioned RF architectures are co-located and mmWave signal propagation has an important feature of multipath sparsity in both the temporal and spatial domains \cite{Pi,T.S.Rappaport}, it is expected that the potentially available diversity and multiplexing gains are not large for the co-located antenna deployment. In order to enlarge diversity/multiplexing gains in mmWave massive MIMO communication systems, this paper considers a more general antenna array architecture, called \emph{distributed antenna subarray architecture}, which includes co-located array architecture as special cases. It is pointed out that, deploying distributed antennas has been shown a promising technique to increase spectral efficiency and expand coverage of wireless communication networks \mbox{\cite{Clark}--\cite{Gimenez}}. As such, it is of great interest to consider distributed antenna deployment in the context of mmWave massive MIMO systems.

The diversity-multiplexing tradeoff (DMT) is a compact and convenient framework to compare different MIMO systems in terms of the two main and related system indicators: data rate and error performance \cite{Zheng, Tse, Yuksel, D.Tse}. This tradeoff was originally characterized in \cite{Zheng} for MIMO communication systems operating over independent and identically distributed (i.i.d.) Rayleigh fading channels. The framework has then ignited a lot of interests in analyzing various communication systems and under different channel models. For a mmWave massive MIMO system, how to quantify the diversity performance and characterize its DMT is a fundamental and open research problem. In particular, to the best of our knowledge, until now there is no unified diversity gain analysis for mmWave massive MIMO systems that is applicable to both co-located and distributed antenna array architectures.

To fill this gap, this paper investigates the diversity performance of mmWave massive MIMO systems with the proposed distributed subarray architecture (the multiplexing performance will be investigated in another paper). The focus is on the asymptotical diversity gain analysis in order to
find out the potential diversity advantage provided by multiple distributed antenna arrays. The obtained analysis can be used conveniently to compare various mmWave massive MIMO systems with different distributed antenna array structures. The main contributions of this paper are summarized as follows: First, for a single-user system employing the proposed distributed subarray architecture, a diversity gain expression is obtained when the number of antennas at each subarray increases without bound. This expression clearly indicates that one can obtain a large diversity gain and/or multiplexing gain by employing the proposed distributed subarray architecture. Second, the diversity gain analysis is extended to the multiuser scenario with downlink and uplink transmission, as well as the single-user system employing the conventional partially-connected RF structure based on the distributed subarrays. Simulation results are provided to corroborate the analysis results and show that the distributed subarray architecture yields significantly better diversity performance than the co-located single-array architecture.

The remainder of this paper is organized as follows. Section II describes the massive MIMO system model and hybrid processing with the distributed subarray architecture in mmWave fading channels. Section III provides the asymptotical diversity analysis for the single-user mmWave system. In Sections IV and V, the diversity gain analysis is extended to the multiuser scenario and the scenario with the conventional partially-connected RF architecture, respectively. Numerical results are presented in Section VI. Section VII concludes the paper.

Throughout this paper, the following notations are used. Boldface upper and lower case letters denote matrices and column vectors, respectively. The superscripts $(\cdot)^T$  and  $(\cdot)^H$ stand for transpose and conjugate-transpose, respectively. $\mathrm{diag}\{a_1,a_2,\ldots,a_N\}$ stands for a diagonal matrix with diagonal elements $\{a_1,a_2,\ldots,a_N\}$. The expectation operator is denoted by $\mathbb{E}(\dot)$.  $[\mA]_{ij}$ gives the $(i,j)$th  entry of matrix $\mA$. $\mA \bigotimes\mB$ is the Kronecker product of $\mA$ and $\mB$. We write a function $a(x)$ of $x$ as $o(x)$ if $\lim_{x \to 0}a(x)/x=0$. Finally, $\mathcal{CN}(0, 1)$  denotes a circularly symmetric complex Gaussian random variable with zero mean and unit variance.

\section{System Model}

\begin{figure*}[t]
\centering
\includegraphics[scale = 0.8]{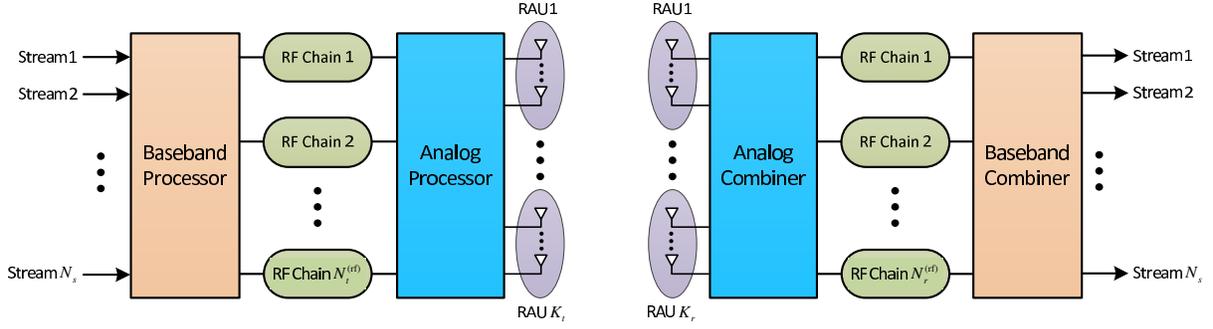}
\caption{Block diagram of a mmWave massive MIMO system with distributed antenna arrays.}
\label{SYS1}
\end{figure*}

\begin{figure}[t]
\centering
\includegraphics[scale = 0.6]{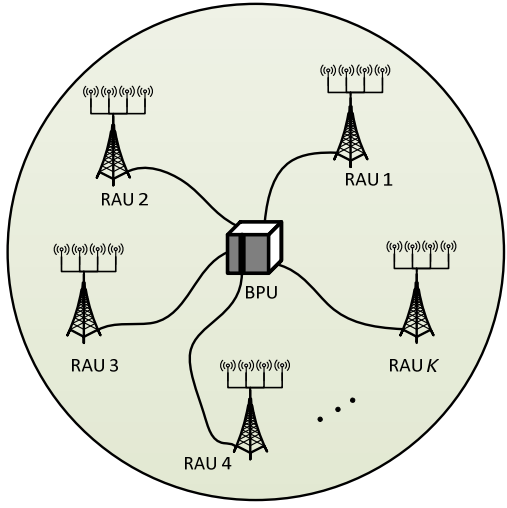}
\caption{Illustration of distributed antenna array deployment.}
\label{SYS2}
\end{figure}

Consider a single-user mmWave massive MIMO system as shown in Fig. \ref{SYS1}. The transmitter is equipped with a distributed antenna array to send $N_s$ data streams to a receiver, which is also equipped with a distributed antenna array. Here, a distributed antenna array means an array consisting of several remote antenna units (RAUs) (i.e., antenna subarrays) that are distributively located, as depicted in Fig. \ref{SYS2}. Specifically, the antenna array at the transmitter consists of $K_t$ RAUs, each of which has $N_t$ antennas and is connected to a baseband processing unit (BPU) by fiber. Likewise, the distributed antenna array at the receiver consists of $K_r$ RAUs, each having $N_r$ antennas and also being connected to a BPU by fibers. Such a MIMO system shall be referred to as a $(K_t, N_t, K_r, N_r)$ distributed MIMO (D-MIMO) system. When $K_t=K_r=1$, the system reduces to a conventional co-located MIMO (C-MIMO) system.

The transmitter accepts as its input $N_s$ data streams and is equipped with $N_t^{({\rm rf})}$ RF chains, where $N_s \leq N_t^{({\rm rf})} \leq N_t K_t $. Given $N_t^{({\rm rf})}$ transmit RF chains, the transmitter can apply a low-dimension  $N_t^{({\rm rf})} \times N_s$ baseband precoder, $\mW_t$, followed by a high-dimension  $K_tN_t \times N_t^{({\rm rf})}$ RF precoder, $\mF_t$. Note that amplitude and phase modifications are feasible for the baseband precoder $\mW_t$, while only phase changes can be made by the RF precoder $\mF_t$ through the use of variable phase shifters and combiners. The transmitted signal vector can be written as
\be \mx = \mF_t\mW_t\ms, \ee
where $\ms$ is the $N_s  \times  1$ symbol vector such that $\mathbb{E}[\ms\ms^H] = \frac{P}{N_s}\mI_{N_s}$. Thus $P$ represents the average total input power. Considering a narrowband block fading channel, the $K_rN_r \times 1$ received signal vector is
\be \label{my} \my =\mH\mF_t\mW_t\ms + \mn  \ee
where $\mH$ is $K_rN_r \times K_tN_t$ channel matrix and $\mn$ is a $K_rN_r \times 1$ vector consisting of i.i.d. $\mathcal{CN}(0, 1)$ noise samples. Throughout this paper, $\mH$ is assumed known to both the transmitter and receiver. Given that $N_r^{({\rm rf})}$ RF chains (where $N_s \leq N_r^{({\rm rf})} \leq N_rK_r$) are used at the receiver to detect the $N_s$ data streams, the processed signal is given by
\be  \label{mz} \mz =\mW_r^H\mF_r^H\mH\mF_t\mW_t\ms + \mW_r^H\mF_r^H\mn  \ee
where $\mF_r$ is the $ K_rN_r \times N_r^{({\rm rf})} $ RF combining matrix, and $\mW_r$ is the $N_r^{({\rm rf})} \times N_s $ baseband combining matrix.

Furthermore, according to the architecture of RAUs at the transmitting and receiving ends, $\mH$ can be written as
\be \label{mH}
    \mH=\left[\begin{array}{lll}
   \sqrt{g_{11}}\mH_{11} & \cdots & \;\;\sqrt{g_{1K_t}}\mH_{1K_t} \\
    \;\;\;\;\vdots & \ddots & \;\;\;\;\;\;\;\;\vdots \\
  \sqrt{g_{K_r1}}\mH_{K_r1} & \cdots & \sqrt{g_{K_rK_t}}\mH_{K_rK_t}
  \end{array}
   \right].
 \ee
In the above expression,  $g_{ij}$ represents the large scale fading effect between the $i$th RAU at the receiver and the $j$th RAU at the transmitter, which is assumed to be constant over many coherence-time intervals. The normalized subchannel matrix $\mH_{ij}$ represents the MIMO channel between the $j$th RAU at the transmitter and the $i$th RAU at the receiver.

A clustered channel model based on the extended Saleh-Valenzuela model is often used in mmWave channel modeling and standardization \cite{Ayach1}, \cite{Li}, \cite{Zhang2} and it is also adopted in this paper. For simplicity of exposition, each scattering cluster is assumed to contribute a single propagation path.\footnote{This assumption can be relaxed to account for clusters with finite angular spreads and the results obtained in this paper can be readily extended for such a case.} Using this model, the subchannel matrix $\mH_{ij}$ is given by
\be \mH_{ij}=\sqrt{\frac{N_tN_r}{L_{ij}}}\sum_{l=1}^{L_{ij}}\alpha_{ij}^l\ma_r(\phi^{rl}_{ij},\theta^{rl}_{ij})\ma_t^H(\phi^{tl}_{ij},\theta^{tl}_{ij}),   \ee
where $L_{ij}$ is the number of propagation paths, $\alpha_{ij}^l$ is the complex gain of the $l$th ray, and $\phi^{rl}_{ij}$ ($\theta^{rl}_{ij}$) and $\phi^{tl}_{ij}$ ($\theta^{tl}_{ij}$) are its random azimuth (elevation) angles of arrival and departure, respectively. Without loss of generality, the complex
gains $\alpha_{ij}^l$ are assumed to be $\mathcal{CN}(0, 1)$. \footnote{The different variances of $\alpha_{ij}^l$ can easily accounted for by absorbing into the large scale fading coefficients $g_{ij}$.} The vectors $\ma_r(\phi^{rl}_{ij},\theta^{rl}_{ij})$ and $\ma_t(\phi^{tl}_{ij},\theta^{tl}_{ij})$ are the normalized receive/transmit array response vectors at the corresponding angles of arrival/departure. For an $N$-element uniform linear array (ULA) , the
array response vector is \be \ma^{\mathrm{ULA}}(\phi)=\frac{1}{\sqrt{N}}\left[1,{\mathrm e}^{j2\pi\frac{d}{\lambda}\sin(\phi)},\ldots,{\mathrm e}^{j2\pi(N-1)\frac{d}{\lambda}\sin(\phi)}\right]^T \ee
where $ \lambda$ is the wavelength of the carrier and $d$ is the inter-element spacing. It is pointed out that the angle
$\theta$ is not included in the argument of $\ma^{\mathrm{ULA}}$ since the response for an ULA is independent of the elevation angle. In contrast, for a uniform planar array (UPA),  which is composed of $N_h$ and $N_v$ antenna elements in the horizontal and vertical directions, respectively, the array response vector is represented by
\be  \ma^{\mathrm{UPA}}(\phi, \theta)=\ma^{\mathrm{ULA}}_h(\phi)\otimes \ma^{\mathrm{ULA}}_v(\theta), \ee
where
\be \ma^{\mathrm{ULA}}_h(\phi)=\frac{1}{\sqrt{N_h}}\left[1,{\mathrm e}^{j2\pi\frac{d_h}{\lambda}\sin(\phi)},\ldots,{\mathrm e}^{j2\pi(N_h-1)\frac{d_h}{\lambda}\sin(\phi)}\right]^T \ee
and
\be \ma^{\mathrm{ULA}}_v(\theta)=\frac{1}{\sqrt{N_v}}\left[1,{\mathrm e}^{j2\pi\frac{d_v}{\lambda}\sin(\theta)},\ldots,{\mathrm e}^{j2\pi(N^v-1)\frac{d_v}{\lambda}\sin(\theta)}\right]^T. \ee

\section{Diversity Gain Analysis}

The most common performance metric of a digital communication system is the error probability, which can be defined either as the probability of symbol error or the probability of bit error (i.e., the bit error rate (BER)). When communicating over a fading channel, errors obviously depend on specific channel realizations. As such, the random behavior of a fading channel needs to be taken into account, leading to the concept of average error probabilities \cite{Ordonez}. Determining exact expressions for the average error probabilities for a digital communication system operating over a certain fading channel is usually tedious and might not give a clear insight about the system behavior. As such, there is a need to characterize the performance of a communication system in a simple and insightful way. A popular approach is to shift the focus from exact performance analysis to asymptotic performance analysis, i.e., analyzing performance at the high signal-to-noise (SNR) region, as done in \cite{Wang}. This is a reasonable approach since the performance of practical interest is in the high SNR region and in such a region, good approximation can be made on the exact analysis.

In the high-SNR region, the average BER function can be approximated in most cases as \cite{Wang} \be \overline{\mathrm{BER}}\approx (G_c\cdot \bar{\gamma})^{-G_d}  \ee
where $G_d$ and $G_c$ are referred to as the diversity and coding gains, respectively, and $\bar{\gamma}$ is the average receive SNR. The diversity
gain determines the slope of the BER curve versus $\bar{\gamma}$ at high SNR in a log-log scale, whereas the coding gain determines how the curve is shifted along the horizontal axis with respect to a benchmark BER curve $\bar{\gamma}^{-G_d}$. This yields a simple parameterized average BER characterization for high SNR, which can provide meaningful insights on the system performance behavior.

In this section, the diversity gain is first examined for a \emph{generalized selection combining}. The main result is then invoked in the diversity analysis of the distributed mmWave massive MIMO system studied in this paper.

\subsection{Diversity Gain of Generalized Selection Combining}

Selection combining (SC) is the most popular low-complexity combining scheme. With selection combining, the receiver estimates the SNRs of all available diversity branches and then select the one with the highest SNR for detection. For \emph{generalized} selection combining (GSC) considered here, the receiver also estimates the SNRs of all available diversity branches. However, instead of selecting the branch with the highest SNR, it selects a branch with the $l$th highest SNR for detection. It is pointed out that, while such a GSC scheme has no practical interest in its own right, its diversity analysis can be used in performance analysis of the mmWave massive MIMO system considered in this paper.

\begin{Lemma} Consider a GSC system with $L$ receive antennas operating over i.i.d. Rayleigh fading channels. If the receiver selects the branch with the $l$th highest SNR for detection then the system achieves diversity gain \be G_d=L-l+1. \ee
\end{Lemma}

{\em Proof:} Let $F(\gamma)$ and $f(\gamma)$ be the probability density function (PDF) and cumulative distribution function (CDF) of the instantaneous SNRs in all branches, respectively. Let $\bar{\gamma}$ denote the average receive SNR of each branch. With Rayleigh fading, it follows from \cite{Wang} that $F(\gamma)$ and $f(\gamma)$ can be written as
\be  F(\gamma)=1-{\mathrm e}^{-\frac{\gamma}{\bar{\gamma}}}=\frac{\gamma}{\bar{\gamma}}+o\lf(\frac{\gamma}{\bar{\gamma}}\ri) \ee
and
\be  f(\gamma)=\frac{1}{\bar{\gamma}}{\mathrm e}^{-\frac{\gamma}{\bar{\gamma}}}=\frac{1}{\bar{\gamma}}-\frac{\gamma}{{\bar{\gamma}}^2}+o\lf(\frac{\gamma}{\bar{\gamma}}\ri). \ee
If the receiver selects the branch with the $l$th highest SNR for detection, then based on the theory of order statistics \cite{David}, the PDF of the instantaneous receive SNR at the receiver, denoted $\gamma_l$, is given by
\bee &&f_{l:L}(\gamma_l)=\frac{L!}{(L-l)!(l-1)!} [F(\gamma_l)]^{L-l}[1-F(\gamma_l)]^{l-1}f(\gamma_l)\nnb \\
&&= \frac{L!}{(L-l)!(l-1)!} \frac{1}{\bar{\gamma}}\lf(\frac{\gamma_l}{\bar{\gamma}}\ri)^{L-l}+o\lf(\lf(\frac{\gamma_l}{\bar{\gamma}}\ri)^{L-l}\ri).\eee
Applying the above PDF in Proposition 1 in \cite{Wang} leads to the desired result. \hfill $\square$

Lemma 1 can be extended to the case of independent but not identically distributed (i.n.i.d.) Rayleigh fading channels and the result is stated in the next lemma.

\begin{Lemma} Suppose that the GSC system with $L$ receive antennas operates over the i.n.i.d. Rayleigh fading channels. If it selects the path with the $l$th highest SNR for detection, then it can achieve diversity gain \be G_d=L-l+1. \ee
\end{Lemma}

{\em Proof:} Let $\bar{\gamma}_{\mathrm{min}}$ and $\bar{\gamma}_{\mathrm{max}}$ denote the maximum and minimum values of the average receive SNRs of all these $L$ diversity paths, respectively. Furthermore, let $\cal A$ and $\cal B$ denote two GSC systems, each equipped with $L$ receive antennas and operating over i.i.d. Rayleigh fading channels such that the average receive SNRs equal to $\bar{\gamma}_{\mathrm{max}}$ and $\bar{\gamma}_{\mathrm{min}}$, respectively. It is known from Lemma 1 that the diversity gains of these two systems are the same and equal to $L-l+1$ if both systems select the branch with the $l$th highest instantaneous SNR for detection. Furthermore, since the GSC system under consideration cannot have better diversity performance than System $\cal A$ and cannot have worse diversity performance than System $\cal B$, it can then be concluded that the i.n.i.d. system must also achieve the diversity gain of $L-l+1$. \hfill $\square$

\subsection{Diversity Gain Analysis of the Distributed mmWave Massive MIMO System}

From the structure and definition of the channel matrix $\mH$ in Section II, there is a total of $L_s=\sum_{i=1}^{K_r}\sum_{j=1}^{K_t}L_{ij}$ propagation paths. Naturally, $\mH$ can be decomposed into a sum of $L_s$ rank-one matrices, each corresponding to one propagation path. Specifically, $\mH$ can be rewritten as
\be \mH=\sum_{i=1}^{K_r}\sum_{j=1}^{K_t}\sum_{l=1}^{L_{ij}}\tilde{\alpha}_{ij}^l\tilde{\ma}_r(\phi^{rl}_{ij},\theta^{rl}_{ij})\tilde{\ma}_t^H(\phi^{tl}_{ij},\theta^{tl}_{ij}),   \ee
where \be  \tilde{\alpha}_{ij}^l =\sqrt{g_{ij}\frac{N_tN_r}{L_{ij}}}\alpha_{ij}^l, \ee  $\tilde{\ma}_r(\phi^{rl}_{ij},\theta^{rl}_{ij})$ is a $K_rN_r \times 1$ vector whose $b$th entry is defined as
\be \label{rrr}[\tilde{\ma}_r(\phi^{rl}_{ij},\theta^{rl}_{ij})]_b=\left\{\begin{array}{ll}
[\ma_r(\phi^{rl}_{ij},\theta^{rl}_{ij})]_{b-(i-1)N_r}, & b\in Q_i^r\\
0, &  b \notin Q_i^r
     \end{array}
     \right. \ee
where $Q_i^r=((i-1)N_r, iN_r]$. And $\tilde{\ma}_t(\phi^{tl}_{ij},\theta^{tl}_{ij})$ is a $K_tN_t \times 1$ vector whose $b$th entry is defined as
\be \label{ttt} [\tilde{\ma}_t(\phi^{tl}_{ij},\theta^{tl}_{ij})]_b=\left\{\begin{array}{ll}
[\ma_t(\phi^{tl}_{ij},\theta^{tl}_{ij})]_{b-(j-1)N_t}, &  b\in Q_j^t\\
0, & b\notin Q_j^t.
     \end{array}
     \right. \ee
where $Q_j^t=((j-1)N_t, jN_t]$.
\begin{Lemma} Suppose that the antenna configurations at all RAUs are either ULA or UPA. Then all $L_s$ vectors $\{\tilde{\ma}_r(\phi^{rl}_{ij},\theta^{rl}_{ij})\}$ are orthogonal to each other when $N_r \to \infty$. Likewise, all $L_s$ vectors $\{\tilde{\ma}_t(\phi^{tl}_{ij},\theta^{tl}_{ij})\}$ are orthogonal to each other when $N_t \to \infty$.
\end{Lemma}

{\em Proof:} It follows immediately from (\ref{rrr}) and (\ref{ttt}) that if $u \neq v$, then vectors $\{\tilde{\ma}_r(\phi^{rl}_{up},\theta^{rl}_{up})\}$ and $\{\tilde{\ma}_r(\phi^{rl}_{vq},\theta^{rl}_{vq})\}$ are orthogonal. On the other hand, when $u=v$ and $p \neq q$, it is known from Lemma 1 and Corollary 2 in \cite{Ayach1} (also see \cite{Chen}) that vectors $\{\tilde{\ma}_r(\phi^{rl}_{up},\theta^{rl}_{up})\}$  and $\{\tilde{\ma}_r(\phi^{rl}_{vq},\theta^{rl}_{vq})\}$ are orthogonal. The proof that $\{\tilde{\ma}_t(\phi^{tl}_{ij},\theta^{tl}_{ij})\}$ is a set of orthogonal vectors can be shown similarly.\hfill $\square$

\begin{Theorem} Suppose that both sets $\{\tilde{\ma}_r(\phi^{rl}_{ij},\theta^{rl}_{ij})\}$ and $\{\tilde{\ma}_t(\phi^{tl}_{ij},\theta^{tl}_{ij})\}$ are orthogonal vector sets when $N_r \to \infty$ and $N_t \to \infty$. Let $N_s \leq L_s$. Then the distributed massive MIMO system with large $N_r$ and $N_t$ can achieve a diversity gain of
\be \label{Gd} G_d=L_s-N_s+1.  \ee

\end{Theorem}
{\em Proof:} The distributed massive MIMO system can be considered as a co-located massive MIMO system with $L_s$ paths that have complex gains $\{\tilde{\alpha}_{ij}^l\}$, receive array response vectors $\{\tilde{\ma}_r(\phi^{rl}_{ij},\theta^{rl}_{ij})\}$ and transmit response vectors $\{\tilde{\ma}_t(\phi^{tl}_{ij},\theta^{tl}_{ij})\}$. Furthermore, order all paths in a decreasing order of the absolute values of the complex gains $\{\tilde{\alpha}_{ij}^l\}$. Then the channel matrix can be written as
\be  \mH=\sum_{l=1}^{L_s}\tilde{\alpha}^l\tilde{\ma}_r(\phi^{rl},\theta^{rl})\tilde{\ma}_t(\phi^{tl},\theta^{tl})^H,  \ee
where $\tilde{\alpha}^1\geq \tilde{\alpha}^2\geq \cdots \geq \tilde{\alpha}^{L_s}$. One can rewrite $\mH$ in a matrix form as
\be  \mH=\mA_r\mD\mA_t^H  \ee
where $\mD$ is a $L_s \times L_s$ diagonal matrix with $[\mD]_{ll}=\tilde{\alpha}^l$, and $\mA_r$ and  $\mA_t$ are defined as follows:
\be  \mA_r=[\tilde{\ma}_r(\phi^{r1},\theta^{r1}),\tilde{\ma}_r(\phi^{r2},\theta^{r2}),\ldots,\tilde{\ma}_r(\phi^{rL_s},\theta^{rL_s})]  \ee
and
      \be  \mA_t=[\tilde{\ma}_t(\phi^{t1},\theta^{t1}),\tilde{\ma}_t(\phi^{t2},\theta^{t2}),\ldots,\tilde{\ma}_t(\phi^{tL_s},\theta^{tL_s})].  \ee
Since both $\{\tilde{\ma}_r(\phi^{rl},\theta^{rl})\}$ and $\{\tilde{\ma}_t(\phi^{tl},\theta^{tl})\}$ are orthogonal vector sets when $N_r \to \infty$ and $N_t \to \infty$, $\mA_r$ and  $\mA_t$  are asymptotically unitary matrices. Then one can form a singular value decomposition (SVD) of matrix $\mH$ as
\be \label{SVD}\mH=\mU\mSigma\mV^H=[\mA_r|\mA_r^{\bot}]\mSigma [\tilde{\mA}_t|\tilde{\mA}_t^{\bot}]^H  \ee
where $\mSigma$ is a diagonal matrix containing all singular values on its diagonal, i.e.,
 \be [\mSigma]_{ll}=\left\{\begin{array}{ll}
 |\tilde{\alpha}^l|, & \mbox{for}\; 1 \leq  l\leq L_s\\
0, & \mbox{for}\; l>L_s
     \end{array}
     \right. \ee
and the submatrix $\tilde{\mA}_t$ is defined as
\be  \tilde{\mA}_t=[{\mathrm e}^{-j\psi_1}\tilde{\ma}_t(\phi^{t1},\theta^{t1}),\ldots,{\mathrm e}^{-j\psi_{L_s}}\tilde{\ma}_t(\phi^{tL_s},\theta^{tL_s})]   \ee
where $\psi_l$ is the phase of complex gain $\tilde{\alpha}^l$ corresponding to the $l$th path.

Based on (\ref{SVD}), the optimal precoder and combiner are chosen, respectively, as
\be  \label{Ft} [\mF_t\mW_t]_\mathrm{opt}=[{\mathrm e}^{-j\psi_1}\tilde{\ma}_t(\phi^{t1},\ldots,{\mathrm e}^{-j\psi_{L_s}}\tilde{\ma}_t(\phi^{tN_s},\theta^{tN_s})]  \ee
and
\be  \label{Fr}[\mF_r\mW_r]_\mathrm{opt}=[\tilde{\ma}_r(\phi^{r1},\ldots,\tilde{\ma}_r(\phi^{rN_s},\theta^{rN_s})].  \ee

To summarize, when $N_t$ and  $N_r$ are large enough, the massive MIMO system can employ the optimal precoder and combiner given in (\ref{Ft}) and (\ref{Fr}), respectively. Then it follows from the above SVD analysis that the instantaneous SNR of the $l$th data stream is given by
\be  \mathrm{SNR}_l=\frac{P}{N_s}|\tilde{\alpha}^l|^2, \;\; l=1,2,\ldots, N_s.        \ee

Now the detection of the $l$th data stream is equivalent to the detection in a generalized selection combining system, which selects the path with the $l$th highest SNR for detection. Therefore, it follows from Lemma 2 that the detection performance of the $l$th data stream has a diversity gain $L_s-l+1$. Since the overall BER is the arithmetic mean of individual BERs, i.e.,\linebreak
$\overline{\mathrm{BER}}=\frac{1}{N_s}\sum_{l=1}^{N_s} \overline{\mathrm{BER}}(l)$,
the system's diversity gain equals to the diversity gain in detecting the $N_s$th data stream, which is the worst among all data streams. Therefore, the result in (\ref{Gd}) is obtained. \hfill $\square$

{\em Remark 1:} When $N_t$ and $N_r$ are large enough, (\ref{SVD}) indicates that the system multiplexing gain is at most equal to $L_s$. This is reasonable since there exist only $L_s$ effective singular values in the channel matrix $\mH$. Theorem 1 provides a simple diversity-multiplexing tradeoff of a mmWave massive MIMO system: adding one data stream to the system decreases the diversity gain by one, whereas removing one data stream increases the diversity gain by one. Such a tradeoff is useful in designing a system to meet requirements on both data rate and error performance.

{\em Remark 2:} {Under the case where $N_t$ and $N_r$ are large enough, it can be deduced from the proof of Theorem 1 that the diversity performance of the mmWave massive MIMO system only depends on the singular value set $\{\tilde{\alpha}^l\}$ and is not influenced by how sub-matrices $\{\sqrt{g_{ij}}\mH_{ij}\}$ are placed in the channel matrix $\mH$ (see further discussion of Fig. \ref{SIM_10} on this point).}

\begin{Corollary} Consider the scenario that the antenna configuration at each RAU is ULA.  Also assume that $L_{ij}=L$ for any $i$ and $j$. Let $N_s \leq K_rK_tL$. When both $N_t$ and $N_r$ are very large, the distributed massive MIMO system can achieve a diversity gain
\be \label{Gd2} G_d=K_rK_tL-N_s+1.  \ee
In particular, when $K_r=K_t=1$, the massive MIMO system with co-located antennas arrays can achieve a diversity gain
\be \label{Gd3} G_d=L-N_s+1  \ee
\end{Corollary}

{\em Remark 3:} Corollary 1 implies that for a mmWave co-located massive MIMO system, its diversity gain and multiplexing gain are limited and at most equal to the number of paths $L$. However, these gains can be increased by employing the distributed antenna architecture and can be scaled up proportionally to $K_rK_t$.

\section{Diversity Gain Analysis with the Conventional Partially-Connected Structure}

\begin{figure*}[t]
\centering
\includegraphics[scale=0.9]{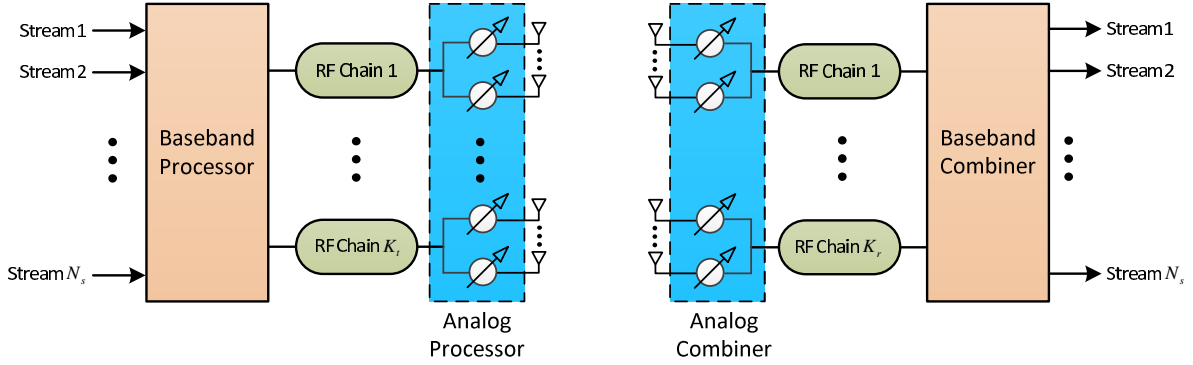}
\caption{Block diagram of a mmWave massive MIMO system with the conventional partially-connected RF architecture.}
\label{SYS3}
\end{figure*}

The previous section has analyzed the diversity gain for the massive MIMO system with the general fully-connected RF architecture.
This section focuses on a massive MIMO system employing the conventional partially-connected RF architecture as illustrated in Fig. \ref{SYS3}. Here the transmitter equipped with $K_t$ RF chains sends $N_s$ data streams to the receiver equipped with $K_r$ RF chains. Each RF chain at the transmitter or receiver is connected to only one RAU. It is assumed that $N_s \leq \min\{K_t,K_r\}$. The numbers of antennas per each RAU at the transmitter and receiver are fixed as $N_t$ and $N_r$, respectively. Note that $N_t \gg N_s$ and $N_r \gg N_s$. Both the transmitter and receiver employ very small digital processors and very large analog processors, represented respectively by $\mW_t$ and $\mF_t$ for the transmitter, and $\mW_r$ and $\mF_r$ for the receiver.

As before, denote by $\ms$ the transmitted symbol vector, by $\mH$ the fading channel matrix, and by $\mn$ the noise vector. Then at the receiver the processed signal vector $\mz$ is given by (\ref{mz}), whereas $\mH$ is described as in (\ref{mH}). Due to the partially-connected RF architecture, the analog processors $\mF_t$ and $\mF_r$ are block diagonal matrices, expressed as
\be  \mF_t=\mathrm{diag}\{\mf_{t1},\mf_{t2},\ldots,\mf_{tK_t} \} \ee
and \be  \mF_r=\mathrm{diag}\{\mf_{r1},\mf_{r2},\ldots,\mf_{rK_r} \} \ee
where $\mf_{ti}$ denotes the $N_t \times 1$ steering vector of phases for the $i$th RAU at the transmitter, and $\mf_{rj}$ the $N_r \times 1$ steering vector of phases for the $j$th RAU at the transmitter.

\begin{Theorem} Consider the case that the antenna array configuration at each RAU is ULA and $L_{ij}=L$ for any $i$ and $j$. In the limit of large $N_t$ and  $N_r$, the distributed massive MIMO system with partially-connected RF architecture can achieve a diversity gain
\be \label{Gdd} G_d=(K_t-N_s+1)(K_r-N_s+1)L.  \ee
\end{Theorem}

{\em Proof:} When $N_t$ and $N_r$ are very large, the diversity gain analysis is similar to that in Theorem 1. For the first data stream that enjoys the best path, it is simple to see that its diversity gain is the largest and equal to $K_rK_tL$. This is because the detection of the first data stream is equivalent to a selection combining system operating with $K_rK_tL$ paths. However, for the second data stream, due to the structure of $\mF_t$ and $\mF_r$, its detection is equivalent to a selection combining system operating with $(K_r-1)(K_t-1)L$ paths. Therefore, it can be inferred that its diversity gain is equal to $(K_r-1)(K_t-1)L$. Similarly, for the last data stream among the $N_s$ data streams, its diversity gain is $(K_r-N_s+1)(K_t-N_s+1)L$. It then follows that the diversity gain of the whole system is just $(K_r-N_s+1)(K_t-N_s+1)L$. \hfill $\square$

{\em Remark 4:} Comparing the diversity gain given in Corollary 1 with that given in Theorem 2 reveals that when $N_s=1$ the diversity gains with the two systems under consideration are the same. However, when $N_s>1$, the proposed distributed antenna system with fully-connected RF architecture achieves a higher diversity gain than the system with the partially-connected architecture, and the gap between the two diversity gains is $(N_s-1)[(K_r+K_t-N_s+1)L-1]$.

\section{Diversity Gain Analysis for the Multiuser Scenario}

\begin{figure*}[t]
\centering
\includegraphics[scale=0.9]{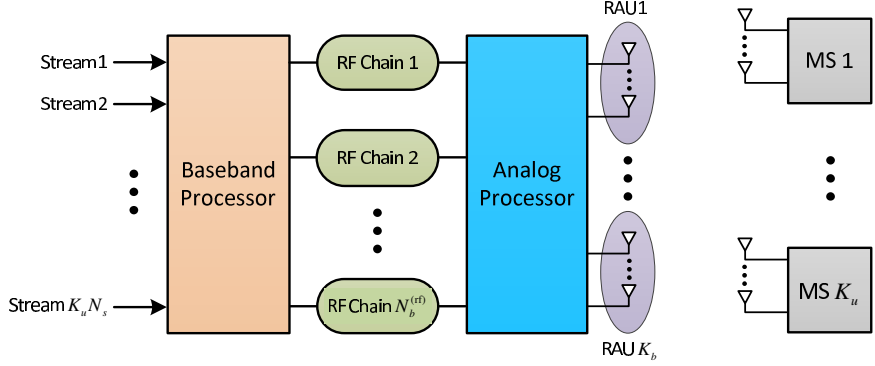}
\caption{Block diagram of a multiuser mmWave system with distributed antenna arrays.}
\label{SYS4}
\end{figure*}

This section considers the downlink communication in a massive multiuser MIMO system as illustrated in Fig. \ref{SYS4}. Here the base station (BS) employs $K_b$ RAUs with each having $N_b$ antennas and $N_b^{({\rm rf})}$ RF chains to transmit data streams to $K_u$ mobile stations. Each mobile station (MS) is equipped with $N_u$ antennas and $N_u^{({\rm rf})}$  RF chains to support the reception of its own $N_s$ data streams. This means that there is a total of $K_uN_s$ data streams transmitted by the BS. The numbers of data streams are constrained as $K_uN_s \leq N_b^{({\rm rf})}\leq K_bN_b$ for the BS, and $N_s \leq N_u^{({\rm rf})}\leq N_u$ for each MS.

At the BS, denote by $\mF_b$ the $K_bN_b \times N_b^{({\rm rf})}$ RF precoder and by $\mW_b$ the $N_b^{({\rm rf})} \times N_sK_u$ baseband precoder. Then under the narrowband flat fading channel model, the received signal vector at the $i$th MS is given by
\be   \my_i=\mH_i \mF_b\mW_b\ms+\mn_i, \; i=1,2,\ldots, K_u \ee
where $\ms$ is the signal vector for all $K_u$ mobile stations, which satisfies $\mathbb{E}[\ms\ms^H] = \frac{P}{K_uN_s}\mI_{K_uN_s}$ and $P$ is the average transmit power. The $N_u \times 1$ vector $\mn_i$ represents additive white Gaussian noise, whereas the $N_u \times K_bN_b$ matrix  $\mH_i$ is the channel matrix corresponding to the $i$th MS, whose entries $\mH_{ij}$ are described as in Section II. Furthermore,
the signal vector after combining can be expressed as
\be   \mz_i=\mW_{ui}^H\mF_{ui}^H\mH_i \mF_b\mW_b\ms+\mW_{ui}^H\mF_{ui}^H\mn_i, \;  i=1,2,\ldots, K_u \ee
where $F_{ui}$ is the $N_u \times N_u^{({\rm rf})}$ RF combining matrix and $\mW_{ui}$ is the $N_u^{({\rm rf})} \times N_s$  baseband combining matrix for the $i$th MS.

\begin{Theorem} Consider the case that all antenna array configurations for the downlink transmission are ULA and $L_{ij}=L$ for any $i$ and $j$ (i.e., all subchannels $\mH_{ij}$ have the same number of propagation paths). In the limit of large $N_b$ and  $N_u$, the downlink transmission in a massive MIMO multiuser system can achieve a diversity gain
\be \label{Gddd} G_d=K_bL-N_s+1.  \ee
\end{Theorem}

{\em Proof:} For the downlink transmission in a massive MIMO multiuser system, the overall equivalent multiuser basedband channel can be written as
\be \label{mHeq}
    \mH_{\rm eq}=\left[\begin{array}{llll}
  \mF_{u1}^H & \mo & \cdots & \;\;\mo \\
   \mo&\mF_{u2}^H  & \cdots & \;\;\mo \\
   \vdots & \vdots & \ddots& \;\;\vdots\\
   \mo& \mo & \cdots & \mF_{uK_u}^H
  \end{array}
   \right]\left[\begin{array}{l}
   \mH_1 \\
   \mH_2\\
    \vdots \\
  \mH_{K_u}
  \end{array}
   \right]\mF_b.
\ee On the other hand, when both $N_b$ and $N_u$ are very large, both receive and transmit array response vector sets, $\{\tilde{\ma}_r(\phi^{rl}_{ij},\theta^{rl}_{ij})\}$ and $\{\tilde{\ma}_t(\phi^{tl}_{ij},\theta^{tl}_{ij})\}$, are asymptotically orthogonal. Therefore the diversity performance for the $i$th user depends only on the subchannel matrix $\mH_i$ and the choices of $\mF_{ui}$ and $\mF_b$. The subchannel matrix $\mH_i$ has a total of $K_bL$ propagation paths. Similar to the proof of Theorem 1, by employing the optimal RF precoder and combiner for the $i$th user, the user can achieve a maximum diversity gain $K_bL-N_s+1$. It is then concluded that the downlink transmission can achieve a diversity gain $G_d=K_bL-N_s+1$. \hfill $\square$

{\em Remark 5:} Theorem 3 implies that when $N_b$ and $N_u$ are large enough,  the available diversity gain $G_d$ does not depend on the number of mobile users $K_u$.

{\em Remark 6:} In a similar fashion, it is easy to prove that the uplink transmission in a massive MIMO multiuser system can also achieve a diversity gain $G_d=K_bL-N_s+1$. Moreover, it can also be proved that when $L=1$,  the system diversity gain is equal to $G_d=K_b$ for the case $N_u=1$, i.e., each MS has only one antenna.

\section{Simulation Results}
\begin{figure}[t]
\centering
\includegraphics[width=3.5 in]{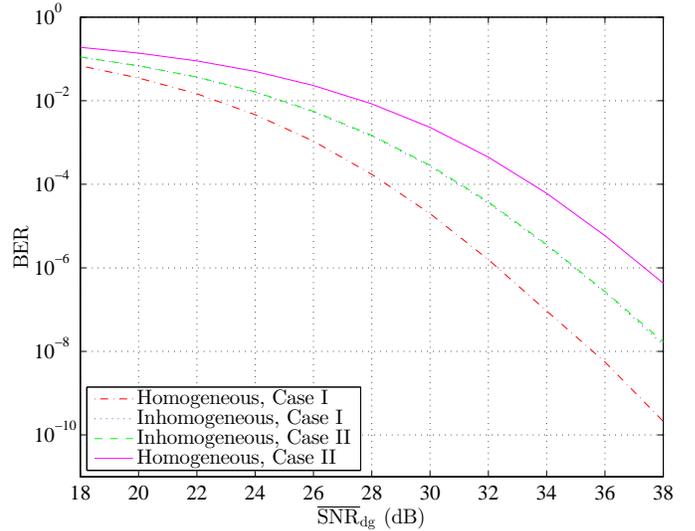}
\caption{BER versus designed SNR: Comparison between homogeneous distributions and inhomogeneous distributions for large scale fading coefficients $\{g_{ij}\}$.}
\label{SIM_10}
\end{figure}

For all simulation results presented in this section, it is assumed that each subchannel matrix $\mH_{ij}$ consists of $L_{ij}=L=3$ paths, each of the large scale fading coefficients $g_{ij}$ equals to $g=-20$ dB (except for Fig. \ref{SIM_10}), and the numbers of transmit and receive RF chains are twice the number of data streams \cite{Sohrabi} (i.e., $N_t^{({\rm rf})}=N_r^{({\rm rf})}=2N_s$). It is further assumed that the variance of AWGN samples is unity and hence the input SNR is the same as the average input power $P/N_s$. For simplicity, only ULA array configuration with $d=0.5$ is considered at RAUs and BPSK modulation is employed for each data stream. With such system configurations, the instantaneous BER is given by $Q(\sqrt{2 \gamma})$ \cite{Goldsmith}, where  $\gamma$ denotes the instantaneous receive SNR and the $Q$-function is defined as $Q(x)=\int_{x}^{\infty} \exp{\left(-\frac{y^2}{2}\right)} {\rm d}y$. For ease of comparison and discussion, introduce the concept of \emph{designed} SNR as
$\overline{\mathrm{SNR}}_{\rm dg}=PN_rN_t/(N_sL)$. This means that $P=\overline{\mathrm{SNR}}_{\rm dg}N_sL/(N_rN_t)$ for a given designed SNR $\overline{\mathrm{SNR}}_{\rm dg}$. In fact, there exists a power scaling law for mmWave communications which states that the data transmit power $P$ can be scaled down proportionally to $1/(N_rN_t)$ to maintain a desirable BER performance \cite{Yue}.

In all simulations, unless stated otherwise, there are three main steps for hybrid digital-analog processing as follows:
\begin{itemize}
\item[(a)] Perform the SVD for channel matrix $\mH$ and find the optimal overall digital precoder and combiner for $N_s$ data streams.

\item[(b)] Form an analog precoder and an analog combiner based on the optimal overall digital precoder and combiner, respectively.

\item[(c)] Perform zero-forcing (ZF) digital detection based on the analog precoder and analog combiner and complete the data detection operation.
\end{itemize}

\begin{figure}[t]
\centering
\includegraphics[width=3.5 in]{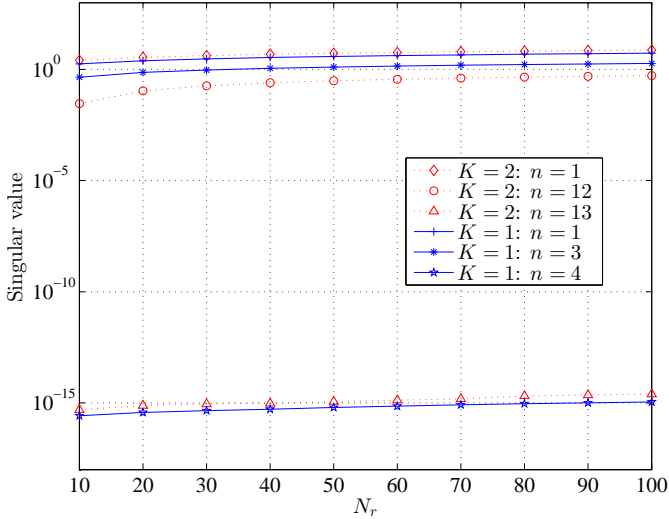}
\caption{Behavior of singular values of channel matrix $\mH$ for $K=1$ and $K=2$.}
\label{SIM_5}
\end{figure}

First, the singular values of channel matrix $\mH$ are studied. Let $K_r=K_t=K$. It is expected that when $N_t$ and $N_r$ are large enough, the number of the effective singular value for the cases $K=2$ and $K=1$ should be equal to $L_s=12$ and $L_s=3$, respectively. To confirm this, Fig. \ref{SIM_5} plots the 1st, the 12th and 13th singular values for $K=2$, and the 1st, the 3th and 4th singular values for $K=1$, when $N_r$ increases from $10$ to $100$,  It can be seen from this figure that as $N_r$ increases, all six singular values slowly increases, but the difference at $N_r=10$ and $N_r=100$ is small. The 13th singular value is very much smaller than the 12th singular value when $K=2$ and it is almost equal to zero. Likewise, the 4th singular value is much smaller than the 3th singular value when $K=1$ and it is almost zero. On the other hand, the 12th singular value under $K=2$ and the 3th singular value under $K=1$ are quite close to their corresponding largest singular values. Thus this figure verifies that the multiplexing gain is in fact at most equal to $L_s$ as stated in Remark 1.

\begin{figure}[t]
\centering
\includegraphics[width=3.5 in]{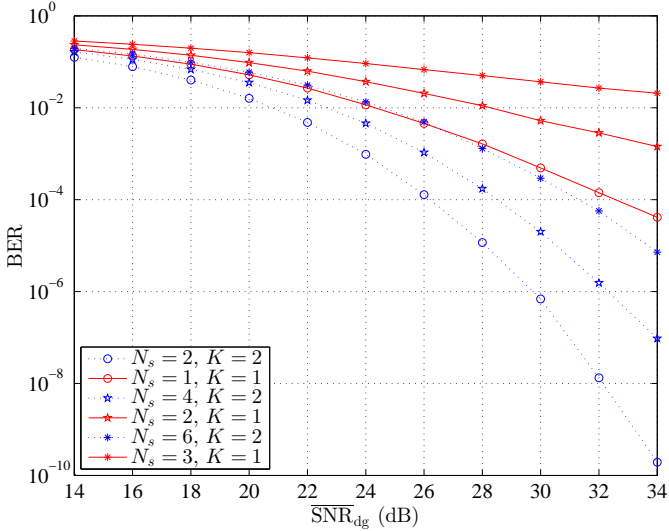}
\caption{BER versus designed SNR: Comparison between distributed and co-located antenna array architectures.}
\label{SIM_6}
\end{figure}

\begin{figure}[t]
\centering
\includegraphics[width=3.5 in]{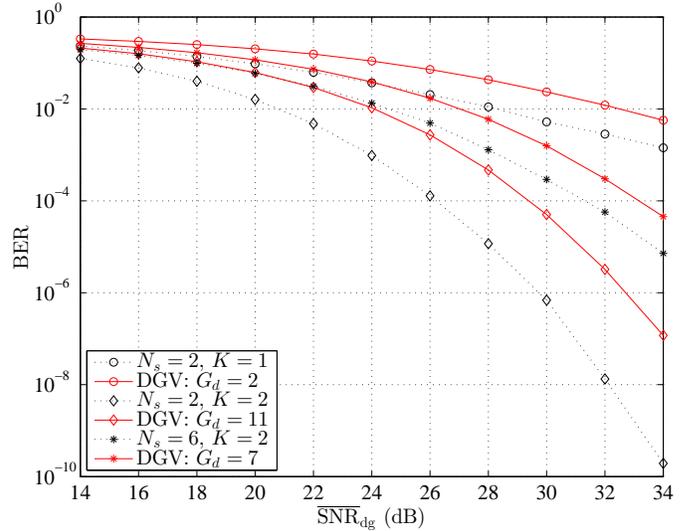}
\caption{BER versus designed SNR: Verifying diversity gain.}
\label{SIM_7}
\end{figure}

Studied next is the diversity performance of a mmWave MIMO system with distributed antenna arrays. With $N_r=N_t=N=50$ and $K_r=K_t=K=2$, Fig. \ref{SIM_6} plots BER curves versus the designed SNR for different numbers of data streams, $N_s=2,4,6$. For comparison, the BER curve obtained in the case of co-located antenna arrays are also plotted for $N_s=1,2,3$. It can be seen that even for the larger number of data streams, the BER performance with distributed antenna arrays is clearly better than that with co-located antenna arrays. Furthermore, as $N_s$ decreases, the BER performance with either distributed or co-located antenna arrays is improved. These observations are expected and agree with Corollary 1, which states that using distributed antenna arrays yields higher diversity gains than using co-located antenna arrays. To verify exactly the diversity gain result given in Corollary 1, Fig. \ref{SIM_7} plots diversity gain verifying (GDV) curves produced by simulating the generalized selection combining (GSC) systems. It can be seen that in the high SNR region, a BER curve with either distributed or co-located antenna arrays has the same slope as the corresponding GDV curve.

\begin{figure}[t]
\centering
\includegraphics[width=3.5 in]{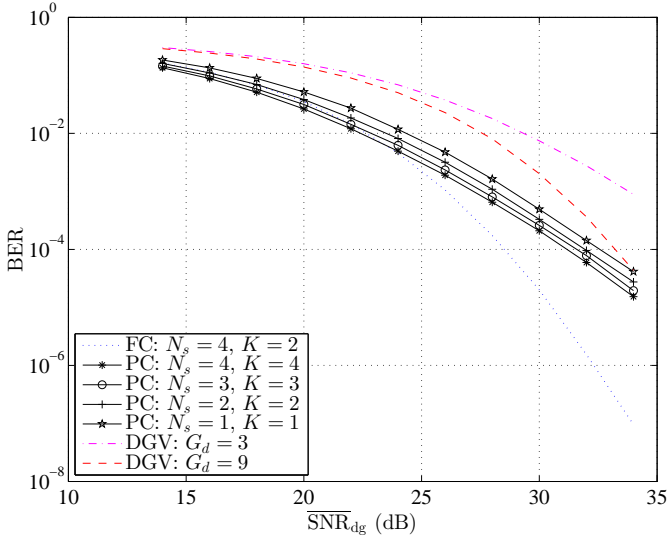}
\caption{BER versus designed SNR: Conventional partially-connected architectures with different numbers of data streams.}
\label{SIM_8}
\end{figure}

Illustrated in Fig. \ref{SIM_8} is the performance with the conventional partially-connected (PC) RF architecture analyzed in Section IV. With this structure, one first carries out the SVDs for subchannel matrices $\{\mH_{ij}\}$ rather than for the whole channel matrix $\mH$ and then forms the analog precoder and analog combiner. Let $K_r=K_t=K$. With $N_r=N_t=N=50$, Fig. \ref{SIM_8} plots the BER curves for the following four cases: $(K=1, N_s=1)$, $(K=2, N_s=2)$, $(K=3, N_s=3)$, and $(K=4, N_s=4)$. It is known from Theorem 2 that the diversity gains for the four cases are identical and equal to $G_d=L=3$. To illustrate this, a DGV curve with diversity gain $G_d=3$ is also plotted in this figure. It can be seen that the system with the conventional PC structure for the four cases can achieve the full diversity gain $3$, while the coding gain increases when both $K$ and $N_s$ increase. For comparison, the BER curve obtained with the general fully-connected (FC) RF structure when $N_s=4$ and $K=2$ is also plotted. The theoretical limit on the diversity gain in this case is $9$, which agrees well with the DGV curve having $G_d=9$. Observe that in the high SNR region the general FC structure yields significantly better diversity performance than the conventional PC structure.

\begin{figure}[t]
\centering
\includegraphics[width=3.5 in]{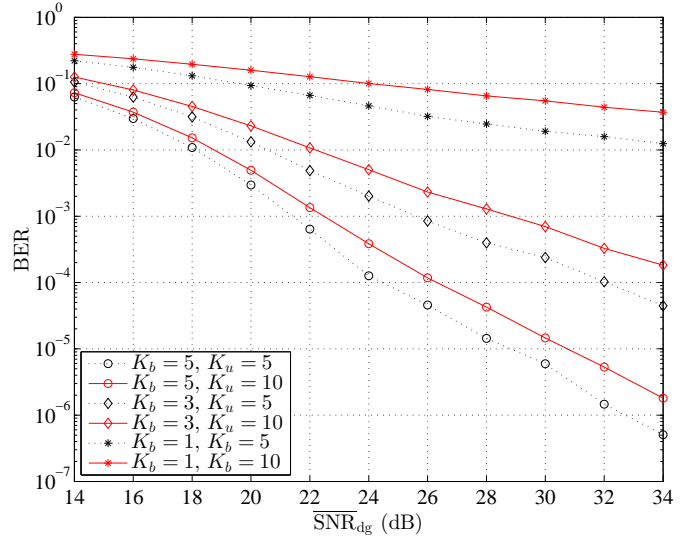}
\caption{BER versus designed SNR: Multiuser scenario with different numbers of subarrays.}
\label{SIM_9}
\end{figure}

Next, when $N_s=1$, we consider the diversity performance with the multiuser downlink scenario where there are 5 or 10 mobile users, each having 10 antennas and each RAU at the BS is equipped with 50 antennas. Due to the fact that there is no cooperation among the users, one first carries out the SVDs for subchannel matrices $\{\mH_i\}$ rather than for the whole channel matrix $\mH$ and then forms the analog precoder for the BS and analog combiners for the users. Note that the BS needs to carry out ZF digital preprocessing before transmitting data. Fig. \ref{SIM_9} plots the BER curves versus the designed SNR for different numbers of subarrays at the BS, namely $K_b=1,3,5$. It can be observed from this figure that as $K_b$ increases, the diversity performance of the multiuser system improves remarkably. This is because, as established in Theorem 3, the diversity gain becomes larger with increasing $K_b$. Furthermore, it can be seen from Fig. \ref{SIM_9} that the system has the same diversity gain for different numbers of users while the coding gain  increases as $K_u$ decreases. This observation agrees with Remark 5.

Finally, the diversity performance of the single-user mmWave massive MIMO system is examined under the scenario that the distributions of large scale fading coefficients, $\{g_{ij}\}$, are inhomogeneous. To this end, let $\mG=[g_{ij}\; ({\rm dB})]$ denote the large scale fading coefficient matrix.  When $N_r=N_t=N=50$ and $K_r=K_t=K=2$, simulation is performed for the following six inhomogeneous $\mG$:
\begin{eqnarray*}
\mG_1=\left[\begin{array}{ll}
   -25 &  -20 \\
  -20 &   -25
  \end{array}
   \right],\;
\mG_2=\left[\begin{array}{ll}
   -20 &  -20 \\
  -25 &   -25
  \end{array}
   \right],\\
\mG_3=\left[\begin{array}{ll}
   -20 &  -25 \\
  -25 &   -20
  \end{array}
   \right],\;
\mG_4=\left[\begin{array}{ll}
   -20 &  -25 \\
  -20 &   -25
  \end{array}
   \right],\\
\mG_5=\left[\begin{array}{ll}
   -25 &  -25 \\
  -20 &   -20
  \end{array}
   \right],\;
\mG_6=\left[\begin{array}{ll}
   -25 &  -20 \\
  -25 &   -20
  \end{array}
   \right].
\end{eqnarray*}
It can be found that the diversity performance for the six inhomogeneous cases are almost the same (see Remark 2). In order to illustrate this interesting phenomenon, Fig. \ref{SIM_10} plots the BER curves versus the designed SNR with $\mG_1$ and $\mG_2$, respectively. For comparison, the two BER curves for the homogeneous distributions with $g=-20\mbox{dB}$ and $g=-25\mbox{dB}$ are also plotted. As expected, the BER curves with the inhomogeneous coefficient distributions are between the two BER curves with homogeneous coefficient distributions. It can be concluded from this figure that the case of inhomogeneous coefficient distributions has the same diversity gain as in the case of homogeneous coefficient distributions.

\section{Conclusions}
This paper has provided asymptotical diversity analysis for mmWave massive MIMO systems with co-located and distributed antenna architectures when the number of antennas at each subarray goes to infinity. Theoretical analysis shows that with a co-located massive antenna array, scaling up the number of antennas of the array can increase the coding gain (array gain), but not the diversity gain. However, if the array is built from distributed subarrays (RAUs), each having a very large number of antennas, then increasing the number of RAUs does increase the diversity gain and/or multiplexing gain. As such, the analysis leads to a novel approach to improve the diversity and multiplexing gains of mmWave massive MIMO systems. It is acknowledged that the asymptotical diversity analysis obtained in this paper is under the idealistic assumption of having perfect CSI. Performing the diversity analysis for mmWave massive MIMO systems under imperfect CSI is important and deserves further research.

\ifCLASSOPTIONcaptionsoff
\newpage
\fi


\end{document}